\documentclass[10pt]{iopart}
\usepackage{color}
\usepackage{graphicx}
\usepackage{latexsym}
\usepackage{amssymb}
\begin{document}
\title[Injection of photoelectrons into dense Ar gas]{Injection of photoelectrons into dense argon gas}
\author{A. F. Borghesani}
\address{Department of Physics, CNISM Unit, University of Padua, via F. Marzolo 8, I-35131 Padua, Italy}
\ead{armandofrancesco.borghesani@unipd.it}
\author{P. Lamp\footnote{Present address: BMW Munich, Germany}}
\address{Max-Planck-Institut f\H ur Physik und Astrophysik, Werner-Heisenberg-Institut f\H ur Physik, F\H ohringer Ring 6, D--8000 M\H unchen 40, Germany}
\ead{peter.lamp@bmw.de}

\begin{abstract}
The injection of photoelectrons in a gaseous or liquid sample is a widespread technique to produce a cold plasma in a weakly--ionized system in order to study the transport properties of electrons in a dense gas or liquid.
We report here the experimental results of of the collection efficiency of photoelectrons injected into dense argon gas at the temperature $T=142.6\,$K as a function of the externally applied electric field and gas density. We show that the experimental data can be interpreted in terms of the so called Young--Bradbury model only if multiple scattering effects due to the dense environment are taken into account when computing the scattering properties and the energetics of the electrons.
\end{abstract}


\section{Introduction}

Injection of electrons from a metal into a gaseous or fluid dielectric is a process of technological relevance whose theoretical understanding is not yet complete \cite{schmidt1997,Allen:1984uq}. Electrons emitted either by photo-- or tunnel cathodes are injected directly into the conduction band of the dielectric medium and, hence, the energy separation of the Fermi level in the metal and the bottom of the conduction band can be measured \cite{Blossey:1974fk}. 

Once injected into a medium, the hot carriers lose energy by means of scattering events that eventually lead them to thermalization. 
Through the same physical mechanisms of scattering, combined with the action of an externally applied electric field and of the image force field, some of the electrons are captured back by the cathode and are not collected. The collection efficiency may thus give useful pieces of 
information on the scattering processes a hot electron undergoes on its way to thermalization \cite{silver1970}.

Rare gases represent a practical realization of disordered systems and the possibility to vary their density 
 in the range between dilute gas $(N\approx 10^{-2}\,$nm$^{-3})$ and liquid $(N\approx 20\,$nm$^{-3})$ offers a unique opportunity for studying
 how electron states and transport  depend on density and degree of disorder.
  
The scattering processes responsible of the drift mobility of electrons are now well understood and can be described up to intermediate densities within a picture in which multiple scattering effects modify the single scattering picture, which is valid at extremely low densities, and suitably dress the electron--atom scattering cross section \cite{BSL}. Whereas for gases of positive scattering length, e.g., helium and neon, this picture breaks down at even higher densities because of electron localization in density fluctuations \cite{hernandez1991}, in gases with negative scattering length, in particular in gaseous and liquid argon, electrons still propagate as quasifree particles with very long mean free paths \cite{lekner1967}. 

In the past, Young and Bradbury developed a theory relating  the actually collected charge to injection energy and applied field \cite{YB}. This model is quite succesful at predicting the overall electric field dependence of the experimental data in a very dilute gas, even in spite of the untenability of some of the assumptions that it is based on.
 
More recently, the injection of electrons in dense argon gas and liquid has been studied by using thin--film cold--cathode emitters \cite{smejtek1973}. The researchers interpreted their experimental data within the YB model even though they acknowledged its flaws. As a result, they were able to detect an unexpected density dependence of the electron--atom momentum--transfer scattering cross section. Unfortunately, their interpretation of the data is spoiled by the unavailability, at that time, of a valid model for the description of scattering at such high densities.

Newer Monte--Carlo (MC) based classical--trajectories numerical simulations  were subsequently carried out in order to statistically study the collection efficiency of the electron injection process in a gas as a function of field and density without the questionable hypotheses of the YB model \cite{kuntz:1136}. As a matter of fact, the results of the numerical simulations agree with the analytical prediction of the YB model as far as the electric field dependence is concerned. However, the simulations do not give any physical explanations for the explicit analytical dependence shown by both the experimental and the simulated data. In addition to that, MC simulations completely fail at predicting the experimentally observed density dependence of the data because they did not take into account, as we will show next, the quantum multiple scattering effects that affect the scattering properties of electrons at high densities.

Thus, within our program of investigating the electron mobility in dense rare gases at very high densities, we decided to study the collection efficiency  of electrons injected into dense argon gas by means of the photoelectric effect  in view of the fact that we now have a well--established theoretical model to describe the behaviour of the quasifree electrons in dense noble gases \cite{BSL}.

 \section{Experimental Details}
 We have used the well-known pulsed photoemission method used in previous electron mobility measurements in neon \cite{borghesani1988,borghesani1990}, helium \cite{Borghesani:2002fk}, and argon \cite{BSL} and we have exploited the same experimental apparatus used for mobility measurements in liquid, gaseous, and critical argon \cite{Eibl:1990fk}. Details of the apparatus have been published elsewhere \cite{Eibl:1990fk,Lamp1989}.
 We recall here only the most relevant features. 
 
 The sample cell consists of a copper block that can withstand gas pressures up to 5 MPa and is contained inside a cryostat for accurate thermoregulation within $1 \,$mK in the temperature range $(100<T<300) \,$K. The cell is filled with the highest--purity (99.9999\% vol.), commercial argon gas. The gas is further purified by circulating it through an Oxisorb filter (Messer Griesheim, Germany) so as to reach the final impurity content of a few tenths of parts per billion required for drift mobility measurements \cite{BSL}. The gas pressure is measured with an accuracy of $\pm 1\,$kPa.
  
 An ultraviolet (UV)--grade quartz window coated with a $\approx 10\,$nm thin  gold layer acts as both photocathode and electrode for the drift voltage \cite{borghesani:2234}. As UV light source a xenon flashlamp (duration $\approx 1\,\mu$s) is used (Hamamatsu, model No. L2435). The spectral distribution of the light emitted by the xenon flashlamp can approximately be described by an asymmetric gaussian peak centered at about $\lambda_{m} =232 \,$nm with left-- and right widths $(+6,\,+28)\,$nm, respectively, corresponding to a photon energy of $\approx 5.4\,$ eV.

 The UV light is guided to the photocathode by means of a UV--grade quartz fiber.  Typically, $\approx 10^{6}$ electrons, i.e., 160 fC, 
 per pulse are released {\em in vacuo}. This amount of charge corresponds to $\approx 0.5\,$ V at the output of the active integrator connected to the anode. In order to improve the signal--to--noise ratio, 256 signals are fetched and averaged together for each value of the electric field applied to the electrodes. The signal waveforms are then analyzed by standard numerical techniques \cite{Borghesani:1990kx}.

  \section{Experimental Results}
  In this paper we report the results for the charge injected into dense argon gas at a temperature $T=142.6\,$K below the critical temperature $T_{c}=150.9\,$K for pressures in the one--phase region $(P< 3.6\, \mathrm{MPa}).$  
  The number density $N$ of the gas is calculated from $T$ and $P$ by means of an accurate equation of state \cite{wagner}.
  
  The applied electric field is in the range $(1<E<400)\,$$\cdot 10^{2}\,$V/m and is small enough to avoid breakdown or gas ionization. The absence of any contact potential effects that might harm the calculated values of $E$ for 
voltages around 1 V is 
confirmed by checking that the drift mobility of electrons  is field-independent for the lowest field strengths \cite{BSL,Borghesani:2001uq}.\

\begin{figure}[!t]
    \begin{center}
\includegraphics[width=\textwidth]{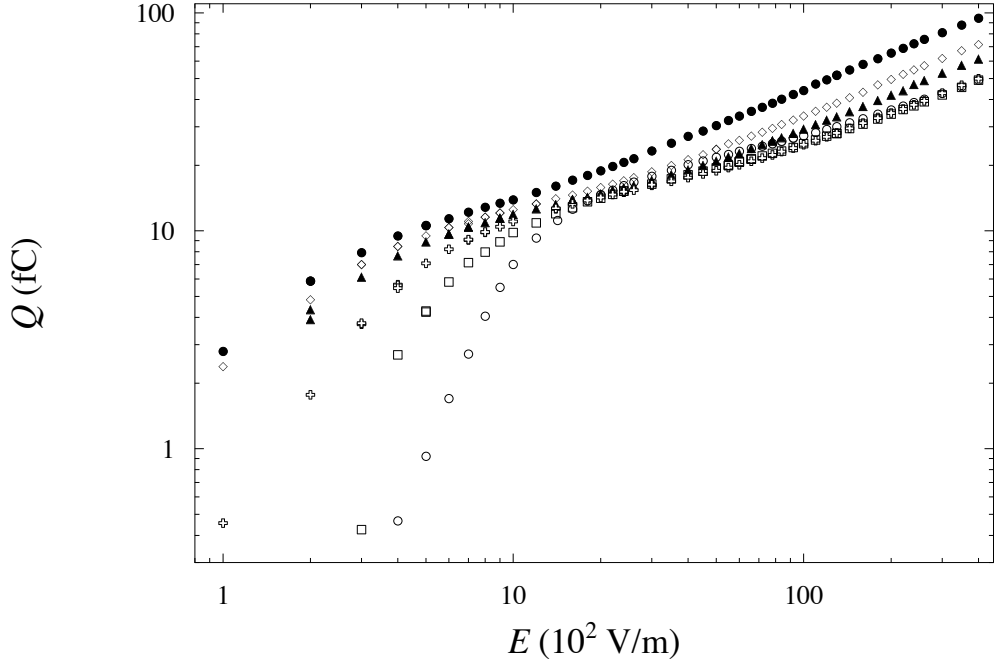}\end{center}
        \caption{\small 
        $Q$ 
        vs $E$ 
        for 
        $N\,(\mathrm{nm}^{-3})\,=0.26$ (closed circles), $
        0.51$ (diamonds),  $
        0.77$ (triangles), $
        1.54$ (crosses), $
        2.33$ (squares), and $
        3.09$ (open circles). 
 \label{fig:QBEIMT143}}
 \end{figure}
In figure \ref{fig:QBEIMT143} we report the experimental results of the charge $Q$ collected by the anode as a function of the applied field $E$ for some densities.  The experimental accuracy is $\approx 10\, \% .$ We do not show all of the measured isopycnal curves  just for the sake of clarity.
 A qualitative analysis of this figure shows that $Q$ steadily increases with increasing $E,$ decreases with increasing $N$ at constant $E,$ does not strongly depend on $N$ for large $E$ when $N$ exceeds some intermediate value, and, finally, shows a change of the dependence on $E$ in a region about $E\approx 15\cdot 10^{2}\,$V/m.

However, such a way to display the results does not help identifying the regularities hidden in the data. Actually, the electric field $E$ is not the best physical variable to describe the data because it has no specific universal significance when the drifting charges do scatter off the gas atoms as in the present case. Electrons are better characterized by the amount of energy gained from the electric field over one mean free path (mfp) $eE\ell = e(E/N\sigma_{\mathrm{mt}}),$ where $\ell = (N\sigma_{\mathrm{mt}})^{-1}$ is the mfp and $\sigma_{\mathrm{mt}}$ is the electron--atom momentum transfer scattering cross section. So, it is customary to plot $Q$ as a function of the density--normalized electric field $E/N$ that is thus proportional, for a given cross section, to the energy in excess of thermal gained from the field during drift.

\begin{figure}[!t]
    \begin{center}
\includegraphics[width=\textwidth]{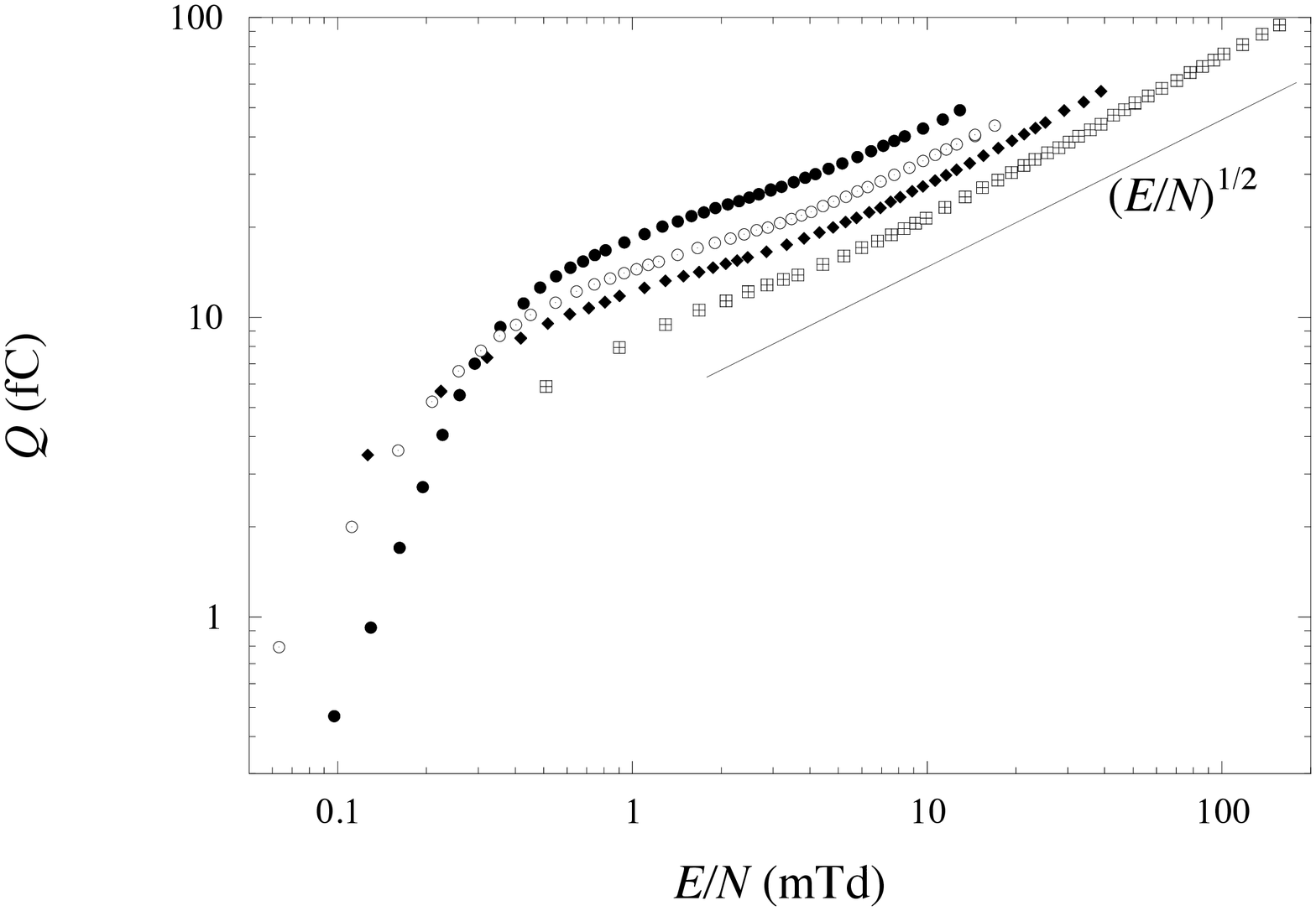}\end{center}
        \caption{\small  $Q$ vs $E/N$ 
        for  $N (\mathrm{nm}^{-3})=3.09$ (closed circles), $
        2.06$ (open circles),  $
        1.03$ (closed diamonds), and $
        0.26$ (squares). 
        $1 \,$mTd$\,=10^{-24}$ V$\,$m$^{2}.$ 
        Solid line: $(E/N)^{1/2}$-law.
 \label{fig:QdiEsuNBEIMT143}}
 \end{figure}
 In figure \ref{fig:QdiEsuNBEIMT143} we plot $Q$ as a function of $E/N$ for some isopycnal curves. These results should be compared with the only other experiment on charge injection in argon gas at a similar temperature though in that experiment hot electrons are injected into the gas by using a tunnel diode as the cathode and though much stronger reduced electric fields are used \cite{smejtek1973}.
 
 \noindent The behaviour of $Q$ as a function of $E/N$ and $N$ is fairly complicated. For $E/N$ well in excess of $\approx 2\,$ mTd $(1\,\mathrm{mTd}=10^{-24} \,\mathrm{V}\,\mathrm{m}^{2})$, $Q\propto (E/N)^{1/2}$ for all densities and the amount of charge collected at constant $E/N$ in this range increases with increasing $N.$ This behaviour compares favorably with the results of Smejtek {\em et al.} \cite{smejtek1973} whose experiment is carried out at $T=160\,$K and only spans the high--field range for $E/N\gtrsim 10\,$mTd. By contrast, the results of the MC simulations \cite{kuntz:1136} show the opposite tendency for the collected charge to decrease with increasing density. Thus, we will next focus on this controversial aspect of the numerical analysis.
 
For $E/N< 2\,$ mTd, the data deviate from the $(E/N)^{1/2}$-law showing a double change of curvature. 

Finally, for $E/N\approx 0.3 \,$mTd, a crossover of the different isopycnal curves takes place so that, at even smaller $E/N,$ the density ordering is reversed with respect to the high--field region.

Owing to the large zero--field electron mobility even at the highest $N,$ $(\mu \approx 0.1\, \mathrm{m}^{2/}\mathrm{Vs})$ \cite{BSL,Borghesani:2001uq}, to the very low impurity content (less than one part per billion O$_{2}$ equivalent) that might give origin to slow O$_{2}^{-} $ ions, and to the small amount of charge injected, we can rule out space--charge effects as the cause of the observed low--field behaviour of the collected charge.

\section{Discussion}
The emission of photoelectrons from a metal cathode into vacuum takes place when the photon energy exceeds the threshold energy, or work function $W_{v},$ of the metal. If emission occurs in a medium, it is found that the threshold energy $W_{m}$ is shifted from its vacuum value by the amount $V_{0},$ which is interpreted as the bottom of the conduction band of the medium \cite{Allen:1984uq}
\begin{equation}
W_{m}=W_{v} +V_{0}
\label{eq:WmWlV0}\end{equation}
 In the case of argon, $V_{0}<0$ \cite{Reininger:1983fk} and less photon energy is required for photoelectron emission than in vacuo. 
 Electrons are then photoemitted into the medium with a broad energy distribution \cite{Fowler:1931vn,DuBridge:1932ys,DuBridge:1933zr} up to  a maximum energy $\mathcal{E}_{0}=\left(hc/\lambda_{m}\right)-W_{m},$ where $\lambda_{m}$ is the shortest wavelength in the flashlamp, $h$ is the Planck's constant and $c$ is the light speed in vacuo.

Once injected into the medium, the epithermal electrons 
drift under the combined influence of diffusion, of their own image field \cite{jackson} that brings them back to the cathode, and of the externally applied electric field $E$ that pulls them toward the anode. 
The net potential energy is given by
\begin{equation}
V(x)= -\frac{1}{4}\frac{e^{2}}{4\pi\epsilon_{0}Kx} - eEx
\label{eq:energiapotenziale}\end{equation}
where $x$ is distance from the cathode and $K$ is the relative dielectric constant of the medium. 
 For the densities of the present experiment $K=1$ within a few percent \cite{maitland}. 

\begin{figure}[!b]
    \begin{center}
\includegraphics[width=\textwidth]{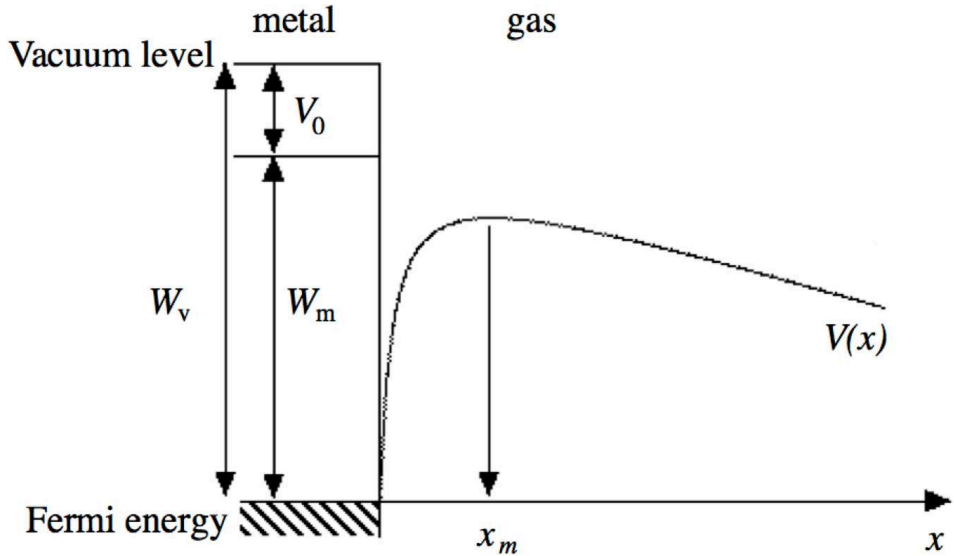}\end{center}
         \caption{\small Schematic diagram of the energy levels at the cathode--gas interface.
         \label{fig:EnLev}}
 \end{figure}

The potential energy $V(x)$ has a maximum at a distance $x_{m}= (e/16\pi\epsilon_{0}KE)^{1/2}$ with value $V_{m}= V(x_{m})= -2eEx_{m}.$ The application of an electric field thus lowers the threshold energy by the Schottky correction $\Delta W = \vert V_{m}\vert.$
It is instructive to evaluate $x_{m}$ and $V_{m}$ in the conditions of the present experiments. By assuming $K\simeq1,$ we get $x_{m}\approx 2\cdot 10^{-6}\,$m and $V_{m}\approx 4\cdot 10^{-5}\,$eV for $E=1\cdot 10^{2}\,$V/m, and $x_{m}\approx 9.5\cdot 10^{-8}\,$m and $V_{m}\approx 8\cdot 10^{-3}\,$eV for $E=4\cdot 10^{5}\,$V/m. 
We note that, at the quite small field strengths of our experiment, the Schottky lowering of the threshold energy is always smaller than the thermal energy $\mathcal{E}_{T}= (3/2) k_{\mathrm{B}}T\approx 18\cdot 10^{-3}\,$eV and the position of the potential energy maximum is located very far from the cathode even on the mfp scale.
 In figure \ref{fig:EnLev} the energy levels at the metal--gas interface are schematically shown.
 
 The emitted electrons are characterized by the initial excess kinetic energy $\mathcal{E}_{0} $ over the barrier. The ultimate fate of an electron injected into the gas, whether it is scattered back to the cathode or is collected by the anode, depends on the distance at which it thermalizes compared to the distance of the potential maximum $x_{m}.$
 
In (pure) argon gas electrons undergo only momentum exchange scattering processes that randomize the electron velocities and lead to a slow loss of their initial kinetic energy. One possibility for the 
electron is to be immediately backscattered well before $x_{m}$ upon injection  and be returned to the cathode \cite{smejtek1973}. The probability that such a backscattered electron might still diffuse forward to the anode over the potential energy maximum or directly tunnel through the potential barrier is negligible owing to the small strength of the electric field, hence the large value of the distance $x_{m},$ and owing to the quite low temperature of the experiment \cite{Blossey:1974fk}.

\noindent Some of the injected electrons may not be backscattered and may slowly lose their excess energy by these elastic scattering processes until they thermalize beyond the potential maximum and are collected by the anode \cite{smejtek1973}. The physical situation, however, is not this simple because the electron escape probability is smaller the higher the initial kinetic energy of the electrons. Actually, an electron that has already crossed the barrier at $x_{m}$ on its way to the anode, and still has sufficient energy to surmount it, may at any time undergo collisions that reverse its motion sending it back across the barrier into the cathode \cite{Allen:1984uq}.

Several attempts have been done in the past at explaining the ratio 
of the observed current to the saturation current, i.e., the current collected in vacuo \cite{thomson,loeb,bekiarian}. The validity of the results obtained so far is difficult to ascertain because it requires an exact solution of the Boltzmann's transport equation which is not yet analytically available unless numerical MC techniques are exploited \cite{kuntz:1136}.

Thermalization is a complicated process in which a huge number of collisions is involved \cite{kuntz:1136,Mozumder:1980uq} and mainly relies on the electron--atom momentum transfer scattering cross section $\sigma_{\mathrm{mt}}.$ 
In spite of this, an oversimplified model due to Young and Bradbury (YB)  \cite{YB} has proven quite succesful at describing the experimental results in low density gases though it has given origin to severe criticism \cite{smejtek1973,kuntz:1136}.

The YB model assumes that the process responsible for the removal of electrons from the current stream is scattering backward at such an angle that the electrons can reach the emitter again. The image field is neglected. The return current is calculated by further assuming only reflection of electrons in their first encounter. In order that an electron returns to the emitter if backscattered at a distance $x$ from it, its kinetic energy $(1/2) mu^{2} $ associated with its velocity $u$ towards the cathode must be greater than the work done by the applied electric field for a displacement over the distance $x,$ $(1/2)mu^{2}\ge eEx.$ 

\begin{figure}[!b]
    \begin{center}
\includegraphics[width=\textwidth]{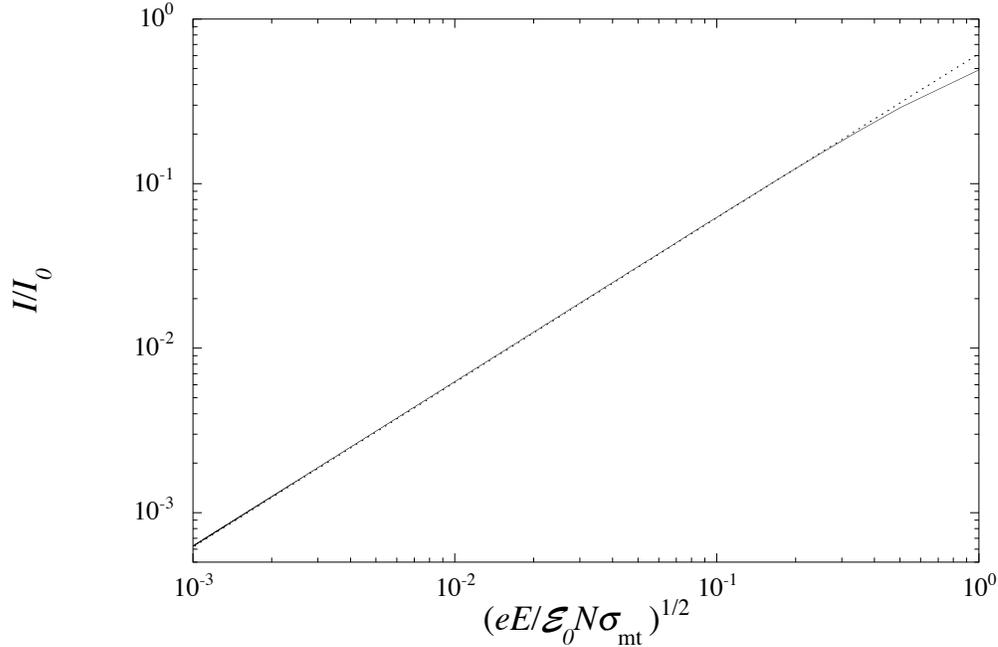}\end{center}
         \caption{\small Ratio of the collected-- to the saturation current in the YB model as a function of the parameter $d=\left(eE/\mathcal{E}_{0}/N\sigma_{\mathrm{mt}}\right)^{1/2}$ (equation \ref{eq:isui0}). Dotted line: equation \ref{eq:YBesun}.\label{fig:isui0YB}}
 \end{figure}

On the other hand, the total kinetic energy at collision with velocity $w$ is $(1/2)mw^{2}= \mathcal{E}_{0}+eEx, $ where $\mathcal{E}_{0}$ is the injection energy.
Electrons return to the cathode if they are scattered within a cone subtending the solid angle
\begin{equation}
\Omega=2\pi \left[ 1- \left(
\frac{x}{x+\mathcal{E}_{0}/eE}\right)^{1/2}\right]
\label{eq:solidangle}\end{equation}
The return probability  $R(x)$ is thus given by $\Omega/4\pi$
\begin{equation}
R(x) =\frac{1}{2} \left[ 1- \left(
\frac{x}{x+\mathcal{E}_{0}/eE}\right)^{1/2}\right]\label{eq:retprob}\end{equation}
By further assuming that the applied electric field is small enough not to significantly deflect the electrons before their first encounter and that $\ell$ is negligible with respect to the distance between cathode and anode, the ratio of the collected current $I$ (or charge $Q$ if current is integrated) to the saturation current $I_{0}$ (or charge $Q_{0}$) can be written as
\begin{eqnarray}
 \frac{I}{I_{0}}\equiv\frac{Q}{Q_{0}}&= &\int\limits_{0}^{\infty}
 \ell^{-1}\exp{\left(-x/\ell\right)} \left[
\frac{x}{x+
\left(\mathcal{E}_{0}/eE\right)
}
 \right]^{1/2}\,\mathrm{d}x\nonumber\\ 
 & =& \int\limits_{0}^{\infty}\e^{-y}\left[
 \frac{y}{y+d^{-2}}\right]^{1/2}
 \,\mathrm{d}y
 \label{eq:isui0}\end{eqnarray}
 where $d^{2}=eE\ell/\mathcal{E}_{0}=\left(eE/\mathcal{E}_{0}N\sigma_{\mathrm{mt}}\right).$
 The integral in equation \ref{eq:isui0} can be evaluated numerically as a function of the parameter $d\propto (E/N)^{1/2}.$ 
 In figure \ref{fig:isui0YB} $I/I_{0}$ is shown as a function of $\left(eE/\mathcal{E}_{0}N\sigma_{\mathrm{mt}}\right)^{1/2}.$ As can be seen, the integral for $d\leq 0.2$ can accurately be approximated by 
\begin{equation}
 \frac{I}{I_{0}}\equiv\frac{Q}{Q_{0}}\simeq 
 A \left(\frac{e}{\mathcal{E}_{0}\sigma_{\mathrm{mt}}}\right)^{1/2}\left( \frac{E}{N}\right)^{1/2}
\label{eq:YBesun}\end{equation}
 where $A$ is a numerical constant of order unity \cite{YB}.

 The condition $d\lesssim 0.2$ sets an upper field strength limit for the validity of equation \ref{eq:YBesun}. In the present experiment we can estimate $\mathcal{E}_{0}\approx 0.35\, $eV, and,  by assuming $\sigma_{\mathrm{mt}}\approx 8\cdot10^{-20}\,$m$^{2}$ for thermal electrons \cite{haddad,Weyhreter:1988kx} (the choice of a more appropriate value of the momentum transfer scattering cross section will accurately be discussed later on), equation \ref{eq:YBesun} is valid for $E/N$ up to $1\,$Td or even more.

Furthermore, it is argued on the basis of plausibility arguments 
that, for equation \ref{eq:isui0} to be valid, the fraction $R$ of electrons that are returned back to the emitter after they have traveled a distance equal to $\ell$ must be smaller than the fraction $T=1-R$ of electrons that are transmitted toward the anode \cite{smejtek1973}. How smaller this fraction has to be is not known. By assuming
\begin{equation} 
T(\ell)-R(\ell)= \left[
\frac{eE\ell}{\mathcal{E}_{0}+eE\ell}
\right]^{1/2}<\alpha 
\label{eq:trasmmaggrifl}\end{equation} 
with $0<\alpha<1,$
for the cases we are interested in, $eE\ell \ll \mathcal{E}_{0},$ i.e., when the energy gained by the electron from the field over a mfp is much smaller than the injection energy, equation \ref{eq:trasmmaggrifl} leads to the condition

\begin{equation}
\left(\frac{E}{N}\right) \gtrsim \alpha^{2}\,
\frac{\mathcal{E}_{0}\sigma_{\mathrm{mt}}}{e}
\label{eq:esunmin}\end{equation}
 If $\alpha =0.1$ as argued in literature \cite{smejtek1973}, and by using the previous estimates for $\mathcal{E}_{0}$ and for $\sigma_{\mathrm{mt}},$ the YB model should be valid for $E/N \gtrsim 0.3\,$Td. This threshold value is far too high as compared with the present experimental data for which the $(E/N)^{1/2}$-law is obeyed for much lower reduced field values. This fact only means that the choice of the numerical value of $\alpha$ is rather arbitrary. Moreover, as it will become clear later, the value of $\sigma_{\mathrm{mt}}$ to be used in equation \ref{eq:esunmin} is not easy to choose owing to the strong energy dependence of the actual cross section and to the presence of multiple scattering effects at high densities.

The data presented in figure \ref{fig:QdiEsuNBEIMT143} cover the low-- to intermediate field range $(5\cdot10^{-5}<E/N<2\cdot10^{-1})\,$Td and partially overlap the previous experiment data that span the higher--field range  $(5\cdot10^{-3}<E/N<4\cdot10^{1})\,$Td \cite{smejtek1973}. 

Our data clearly show that the YB law, equation \ref{eq:YBesun}, reasonably well describes the collected charge data for $E/N\gtrsim 1\,$mTd up to $E/N\approx 0.2\,$Td for all the investigated densities from $N=0.26\,$nm$^{-3}$ up to $N=3.09\,$nm$^{-3}.$ This result extends the validity of the YB model to $E/N$ values one order of magnitude smaller than the previous experiment \cite{smejtek1973} and suggests that the assumption expressed by equation \ref{eq:trasmmaggrifl} is unrealistic.

A detailed inspection of figure \ref{fig:QdiEsuNBEIMT143} further shows that the experimental data do not strictly obey the YB law even at the highest field strength. We believe that the upward deviations from it occur as a consequence of the energy dependence of the cross section that increases with energy for energies above the Ramsauer--Townsend (RT) minimum.

If the YB law, equation \ref{eq:YBesun}, holds true (at least, approximately), the value of the momentum transfer scattering cross section can be deduced at each density $N.$ 
In order to do this, it is better to recast equation \ref{eq:YBesun} in the following form:
\begin{equation}
 \frac{Q}{Q_{0}}=A^{\prime}S(N) \left(\frac{E}{N}\right)^{1/2}
\label{eq:YBesunrecast}\end{equation}
in which $A^{\prime}= Ae^{1/2}$ and $S(N)=\left(\mathcal{E}_{0}\sigma_{\mathrm{mt}}\right)^{-1/2}
.$
The density dependent cross section is then calculated as
\begin{equation}
\sigma
(N)= \frac{S^{2}\left(N_{1}\right)}{S^{2}\left(N\right)}\frac{\mathcal{E}_{0}\left(N_{1}\right)}{\mathcal{E}_{0}\left(N\right)}\sigma_{0}
\label{eq:sigmasme}\end{equation}
where $N_{1}$ is a (low) density taken as a reference, $\sigma_{0}$
 is  a scattering cross section value obtained from low--density gas swarm experiments, and $\mathcal{E}_{0}$ is the injection energy.
 This simple procedure yields a density--dependent cross section, as obtained by Smejtek {\sl et al.} \cite{smejtek1973}, indeed. For reasons that will become clear later, we have simply termed $\sigma(N)$ the cross section determined in this way, rather than using the previous symbol $\sigma_{\mathrm{mt}}.$

 In order to determine $\sigma(N)$ from our data we have first fitted the $Q$ data to the $(E/N)^{1/2}$-law for $E/N\gtrsim 1\,$mTd so as to calculate the slope $S(N).$ The injection energy for $N=0$ in our case is estimated to be $\mathcal{E}_{0}\left(N=0\right)\equiv\mathcal{E}_{0}(0)\approx 0.35\,$eV. The variation of  $\mathcal{E}_{0}$ with $N$ is accounted for by the density dependence of the energy at the bottom of the conduction band of the gas $V_{0}(N) $ 
 \begin{equation}
 \mathcal{E}_{0} (N)= \mathcal{E}_{0}(0) - V_{0}(N)
 \label{eq:e0n}
 \end{equation}
We have used for $V_{0}$ the experimental data of Reininger {\sl et al.}, which are well described by the interpolation formula \cite{Reininger:1983fk}
\begin{equation}
V_{0}(N)= V_{0}\left(N_{0}\right)+ a\left(N-N_{0}\right) +\left(\frac{b}{c}\right)\ln{\left\{\cosh{\left[c\left(N-N_{0}\right)\right]}\right\}}
 \label{eq:v0}\end{equation}
 with $N_{0}=11.03\,$nm$^{-3},$ $V_{0}\left(N_{0}\right)=-0.253\,$eV, $a=- 3.34\cdot 10^{-3}\,$eV$\,$nm$^{-3},$ 
 $b= 2.48\cdot 10^{-2}\,$eV$\,$nm$^{-3},$ and $c=- 0.3\,$nm$^{-3}.$ The interpolation formula is corrected for impurity effects at low density \cite{borghesani1991}.
 
 In our conditions the effect of the density change of $V_{0}$ is quite important. At the highest density of our experiment $N\approx 3.09\,$nm$^{-3},$ the contribution of $V_{0}$ amounts to $\approx 25\, \%$ of $\mathcal{E}_{0}(0).$
  
The normalization constant has been chosen $\sigma_{0}= 3.4\cdot10^{-20}\,$m$^{2}  $ for $N_{1}=0.26\,$nm$^{-3},$ which is consistent with the 
mobility data published elsewhere \cite{BSL,Borghesani:2001uq}.

\begin{figure}[!b]
    \begin{center}
\includegraphics[width=\textwidth]{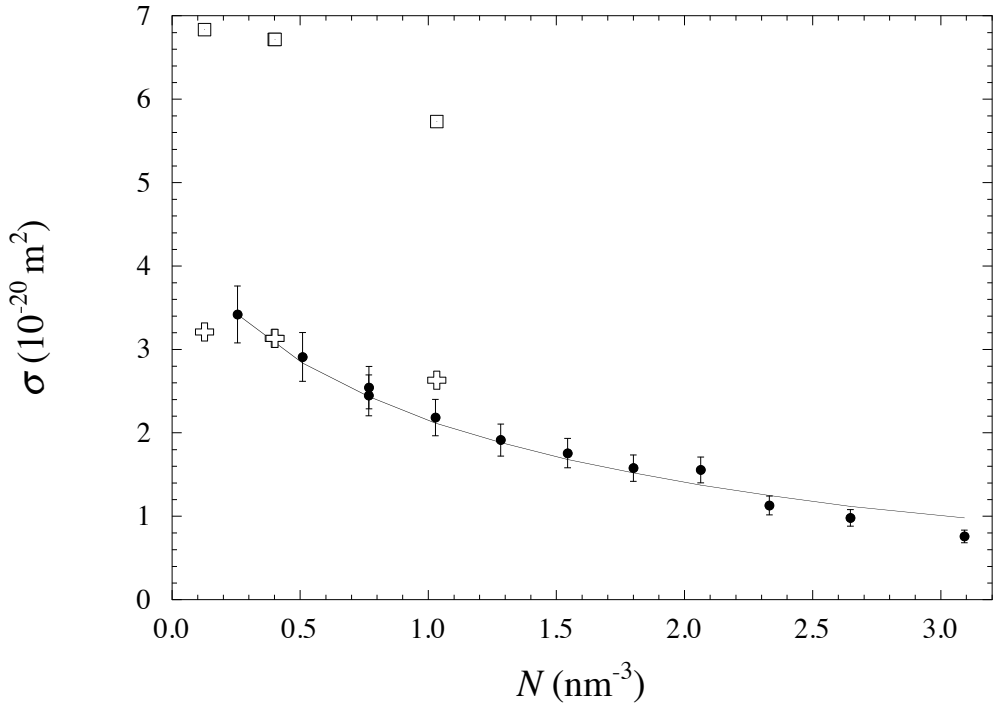}\end{center}
         \caption[Experimentally detected cross section]{\small Density dependence of the momentum transfer scattering cross section determined by equation \ref{eq:sigmasme} at fairly high fields. Dots: present determination of $\sigma$ normalized by the cross section value at $N=0.26$ nm$^{-3}.$ Solid line: $\sigma_{\mathrm{mt}}$ evaluated at the shifted thermal energy (see text). Open squares: data by Smejtek {\sl et al.} \cite{smejtek1973}. Crosses: corrected Smejtek's data (see text).\label{fig:sigmaq2}}
 \end{figure}

In figure \ref{fig:sigmaq2} we plot the density--dependent momentum transfer scattering cross section determined according to the above mentioned procedure. The data of Smejtek {\sl et al.} are also shown for the sake of comparison. 
 
 The observed density dependence of the electron--atom momentum transfer scattering cross section can be easily explained in terms of the model developed by us in order to explain the anomalous density effects of the electron mobility in dense noble gases \cite{borghesani1988,BSL}. For a detailed description of this model we refer to a previous paper \cite{BSL}. We recall here its main features to the specific goal of interpreting the present data.
 
 At the densities of the present experiment the electron de Broglie wavelength, its mfp, and the average interatomic distance become comparable with each other so that multiple scattering effects set in. In particular, the ground state energy of a quasifree electron immersed in a medium is increased with respect to its thermal value by a density--dependent quantum shift \cite{Fermi:1934uq} that is recognized as the bottom of the conduction band $V_{0}(N).$  $V_{0}(N)$ can be written as the sum of potential and kinetic contributions \cite{Springett:1968fk}
 \begin{equation}
V_{0}(N) = U_{P}(N) + E_{K}(N)
 \label{eq:sjc}\end{equation}
$U_{P}(N)<0$  is a potential energy contribution that arises from the screened polarization interaction of the electron with the surrounding atoms whereas $E_{K}(N)$ is a kinetic energy term that is due to excluded volume effects because the volume accessible to the electron shrinks as $N$ increases. It turns out that $E_{K}(N)>0$ and increases with increasing $N.$ $E_{K}$ can be calculated by enforcing the condition that the electron ground state wave function is endowed with average translational symmetry about the equivalent Wigner--Seitz (WS) cell centered about each atom of the gas \cite{Hernandez:1991kx}. This condition leads to the eigenvalue equation
 \begin{equation}
 \tan{\left[ k_{0} \left( r_{s} -\tilde a\left( k_{0}\right)\right)\right]} -k_{0}r_{s}=0
 \label{eq:ews}\end{equation}
that must be solved for the wavevector $k_{0}(N)$ in a selfconsistent way. $r_{s} =\left(3/4\pi N\right)^{1/3}$ is the radius of the WS cell, $\tilde a =\left(\sigma_{t}/4\pi\right)^{1/2}$ is the hard--core radius of the Hartree--Fock potential for rare gas atoms \cite{Springett:1968fk}, and $\sigma_{t}$ is the electron--atom total scattering cross section. 
 Finally, $E_{K}$ is given by
 \begin{equation}
E_{K}= \frac{\hbar^{2}k^{2}_{0}}{2m}
\label{eq:ekws}\end{equation}
  \begin{figure}[!b]
    \begin{center}
\includegraphics[width=\textwidth]{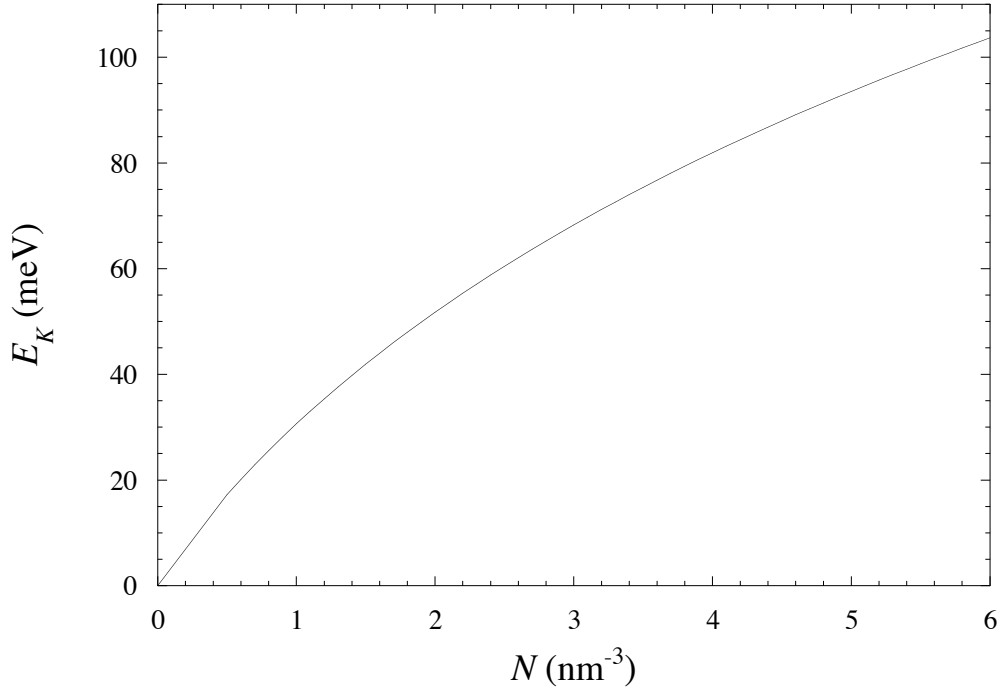}\end{center}
         \caption{\small Density dependence of the kinetic energy shift $E_{K} $ \cite{BSL}.\label{fig:ekn}}
 \end{figure}

\noindent In figure \ref{fig:ekn} we show $E_{K}(N)$ as a function of the gas density calculated according to equation \ref{eq:ekws} by using the cross section reported by Weyrehter {\sl et al.} \cite{Weyhreter:1988kx}.

\noindent The experiments on electron mobility in dense rare gases \cite{borghesani1988,borghesani1990,BSL,Eibl:1990fk,Borghesani:2001uq,Lamp:1994fk} have clearly shown that only the kinetic contribution $E_{K}$ of the total energy shift $V_{0}$ enhances the kinetic energy of electrons during collisions. 
\noindent In this way, the scattering properties of electrons, namely their scattering cross sections, have to be evaluated at the shifted kinetic energy $\mathcal{E}+E_{K}(N).$ In other words, the bottom of the electron energy distribution function is shifted by an amount equal to $E_{K}(N).$

\begin{figure}[!b]
     \begin{center}
\includegraphics[width=\textwidth]{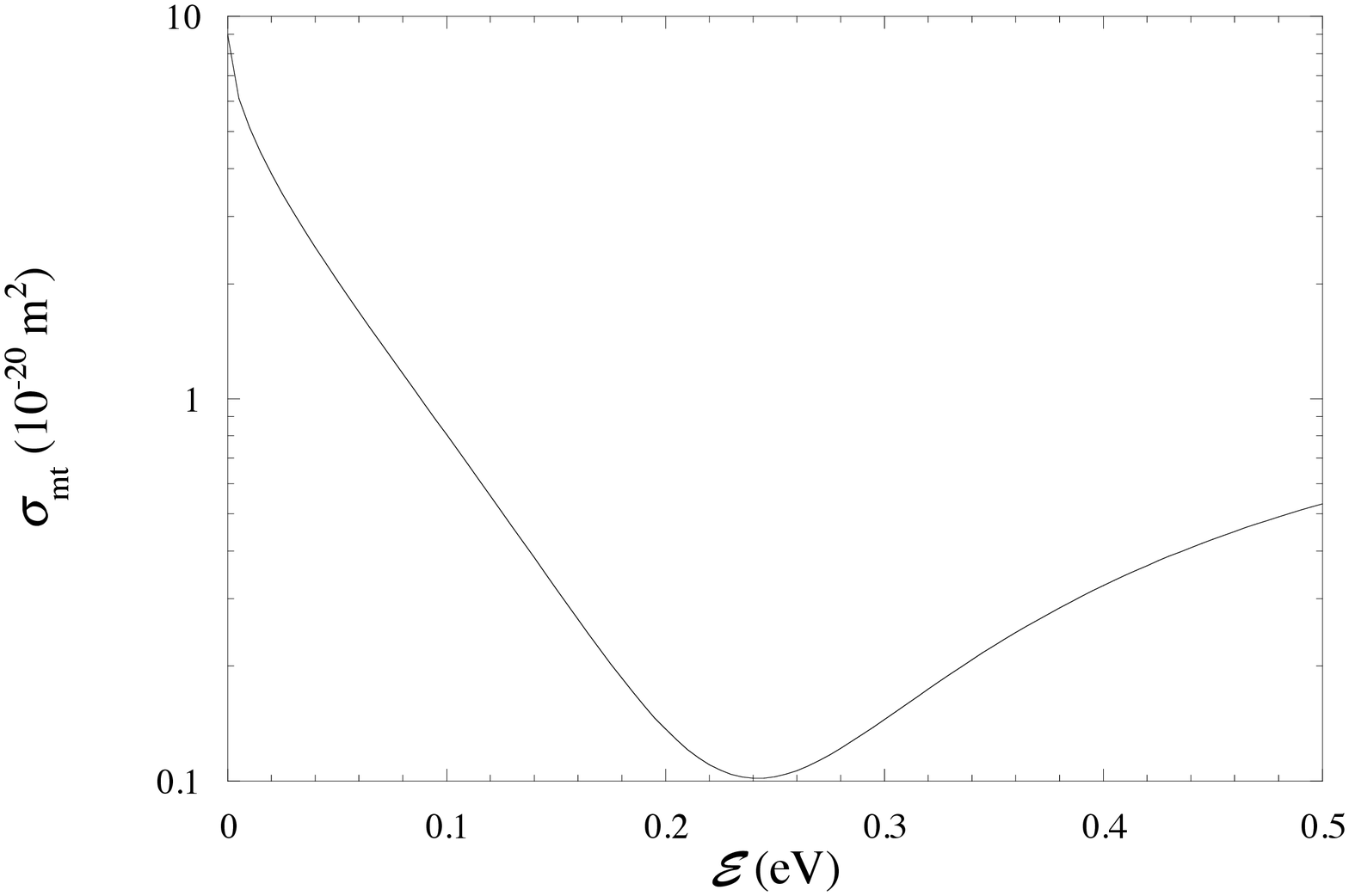}\end{center}
         \caption{\small Energy dependence of the momentum transfer scattering cross section $\sigma_{\mathrm{mt}}$ of argon \cite{Weyhreter:1988kx}. \label{fig:sigmamtAr}}
 \end{figure}

 The dependence of the cross section on the electron energy, shown in figure \ref{fig:sigmamtAr}, and the energy shift by $E_{K}(N)$ are the main physical effects leading to a density dependence of the cross section. 
 Smejtek {\sl et al.}  \cite{smejtek1973} were not able to explain the causes of the observed density dependence of the cross section because the physical picture described above was yet to emerge at that time.
 
 Two more multiple scattering effects come into play when the density is large enough for the electron mfp and wavelength to become comparable. The first one is a quantum self--interference of the electron wave function scattered off atoms located along paths connected by time--reversal symmetry \cite{Ascarelli:1986uq} that leads to an increase of the rate of back--scattering \cite{Atrazhev:1977kx}. 
 
 The second effect is due to correlations among scatterers. The electron  wave packet extends over a wide region encompassing many atoms. The total scattered  wave packet is obtained by coherently summing up all partial scattering 
 amplitudes contributed by each atom and the resulting cross section is enhanced by the static structure factor of the gas \cite{Lekner:1968vn}.
 
The latter two multiple scattering effects deeply influence the propagation of the wave packet and, hence, the electron mobility, indeed, though they do not alter very much the electron energy distribution function. We can neglect them for the analysis of the results of the present experiment mainly because the YB model does not deal with the electron wave packet propagating through the gas from the cathode to the anode but it only treats the charge backscattered in the first encounter and the collected charge is obtained only as a difference between the injected and backreflected charge.

The electron energy distribution function $g\left(\mathcal{E}\right)$ is given by the Davydov--Pidduck distribution function \cite{Wannier,cohen}
\begin{equation}
g\left(\mathcal{E}\right) =C \left\{ - \int\limits_{0}^{\mathcal{E}} \left[
k_{\mathrm{B}}T+ \frac{M}{6mz}\left(
\frac{eE}{N\sigma_{\mathrm{mt}}(z+E_{K}(N))}
\right)^{2}
\right]^{-1}\,\mathrm{d}z
\right\}
\label{eq:dp}\end{equation}
 Here, $k_{\mathrm{B}}$ is the Boltzmann's constant, $M$ and $m$ are the masses of the argon atom and of the electron, respectively. The normalization constant $C$ is such that $\int_{0}^{\infty} z^{1/2}g(z) \,\mathrm{d}z =1.$

In equation \ref{eq:dp} we have explicitly put into evidence that the cross section is evaluated at the shifted energy whereas we have dropped the corrections due to correlation and self--interference effects with respect to the formulas used for the mobility \cite{BSL}.

Once the distribution is known, averages can be calculated. In particular, it can be shown that the electron mean energy remains approximately thermal, except for the contribution $E_{K},$ up to quite high field strengths $E/N\approx 2$ or $3\,$mTd, depending on the density.
We believe that this fact further means that thermalization in dense argon gas must be a fairly rapid process. 

\begin{figure}[!t]
    \begin{center}
\includegraphics[width=\textwidth]{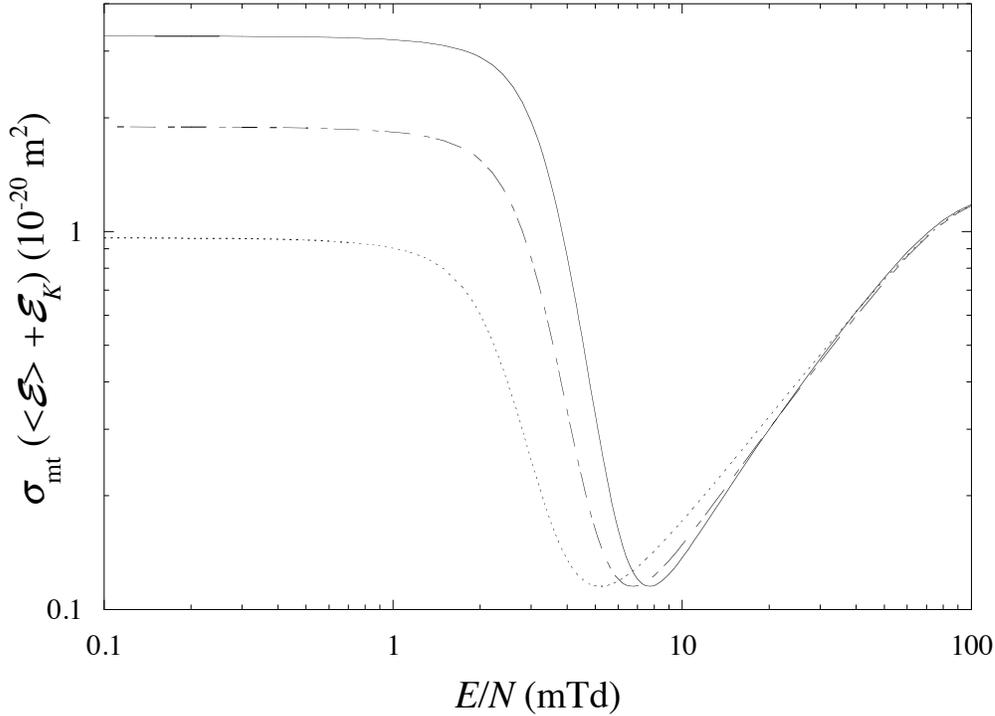}\end{center}
        \caption{\small $E/N-$dependence of the momentum transfer scattering cross section $\sigma_{\mathrm{mt}}$ evaluated at the shifted mean energy $\bar\mathcal{E} = \langle\mathcal{E} \rangle+{E}_{K}$ for $N(\mathrm{nm}^{-3})=0.26$ (solid line), $1.28$ (dash--dotted line), and $3.09$ (dashed line). \label{fig:Sigmamtthermalplusek}}
 \end{figure}

In figure \ref{fig:Sigmamtthermalplusek} we show  the momentum transfer scattering cross section evaluated at the mean shifted energy $\bar \mathcal{E}=\langle\mathcal{E}\rangle +E_{K} ,$ where $\langle \ldots\rangle $ indicates a thermal average. 
As anticipated, the scattering cross section now shows a strong density dependence, especially at low $E/N,$ that is acquired by the effect of the energy shift $E_{K}(N)$ combined with the very rapid decrease of $\sigma_{\mathrm{mt}}$ with energy as shown in figure \ref{fig:sigmamtAr}.
  In the low--field region, electrons are thermal but their mean energy is $\bar\mathcal{E}= (3/2) k_{\mathrm{B}}T + E_{K}(N)>(3/2) k_{\mathrm{B}}T.$ Thus, the average cross section, which can be well approximated in the low--field region by the cross section evaluated at the mean energy $\langle \sigma_{\mathrm{mt}}\left(\mathcal{E}\right)\rangle\simeq \sigma_{\mathrm{mt}}\left(\bar \mathcal{E}\right),$ turns out to be much smaller than if it were evaluated at thermal energy only. For instance, for $T=142.6\,$K  and for $N=0.26 \,$nm$^{-3},$ $(3/2) k_{\mathrm{B}}T \approx 18\,$meV and $E_{K}\approx 10 \,$meV. $\sigma_{\mathrm{mt}}\left(\mathcal{E}=(3/2)k_{\mathrm{B}}T \right) \approx 4\cdot10^{-20}	\,$m$^{2},$ whereas $\sigma_{\mathrm{mt}}\left(\mathcal{E}=\bar\mathcal{E} 
  \right) 
  \approx 3\cdot10^{-20}	\,$m$^{2}.$

In figure \ref{fig:sigmaq2} the solid line represents $\sigma_{\mathrm{mt}}\left(\bar\mathcal{E}\right)$ that compares very favorably with the values of the density dependent cross section obtained by analyzing the collected charge data according to the YB model. This good agreement between theory and experimental data lends credibility to our analysis.

Moreover, we are now in a position to explain the discrepancy between our data and those of Smejtek {\sl et al.} \cite{smejtek1973} (open squares in figure \ref{fig:sigmaq2}) in the density region where comparison is possible. First of all, in the analysis of their data, Smejtek {\sl et al.} assumed the injection energy $\mathcal{E}_{0}\approx 1\,$eV independent of density. So, they disregarded the density variation of the bottom of the conduction band $V_{0}$ by setting ${\mathcal{E}_{0}\left(N_{1}\right)}/{\mathcal{E}_{0}\left(N\right)}=1$ in equation \ref{eq:YBesunrecast}. 
As a consequence of the rapid variation of $V_{0}$ with $N$ they overestimated the cross section by a factor $\left[\mathcal{E}_{0}-V_{0}(N)\right]/\mathcal{E}_{0}$ leading to a correction 
that can be as large as 30 \% for $N=10\,$nm$^{-3}$ if $\mathcal{E}_{0}=1\,$eV. Were the injection energy $\mathcal{E}_{0}<1\, $eV, the overestimation factor would be even worse.  

A second issue is that they normalized the data by using $\sigma_{0}$ values derived from old swarm experiments at very high reduced electric fields $E/N$ \cite{grunberg,Allen:1970ys}. At such high fields, the electron drift mobility does no longer depend on density and it is much smaller than at low fields, thus leading to an estimation of the cross section that is erroneously too large. The old swarm experiments are superseded by more recent ones \cite{Bartels:1973zr,BSL}, in which the drift mobility has been measured also in the limit of low fields. These new experiments have clarified the physical nature of the density dependence of the cross section for measurements in gas under pressure and their results about the cross section are consistent with its determination from low--density swarm data \cite{haddad} and from beam experiment \cite{Weyhreter:1988kx}. The new mobility data \cite{Bartels:1973zr,BSL} allow a much more accurate estimate of $\sigma_{0}$ at the density of Smejtek's experiment.

By performing the corrections relative to the overestimation of the ${\mathcal{E}_{0}\left(N_{1}\right)}/{\mathcal{E}_{0}\left(N\right)}$ factor and  of $\sigma_{0},$ the data of Smejtek {\sl et al.} can now be shown to be in far better agreement with the cross section values obtained in the present experiment, as shown by the crosses in figure \ref{fig:sigmaq2}.

As already stated, the present data cover a range of reduced electric fields that are a couple of order of magnitudes smaller than in the previous experiment \cite{smejtek1973}. In this range quasifree electrons are thermal. It is evident from figure \ref{fig:QdiEsuNBEIMT143} that, as $E/N$ is reduced, for each density there is a crossover region leading to deviations from the YB $(E/N)^{1/2}$-law. In the low field region the collected charge shows a much stronger dependence on $E/N$ than at high fields. Moreover, the higher the density the stronger the field dependence. We do not have any explanations for this behaviour but we propose an analysis to show that it is still related to the properties of the momentum transfer scattering cross section even at quite low reduced fields.

\begin{figure}[!b]
    \begin{center}
\includegraphics[width=\textwidth]{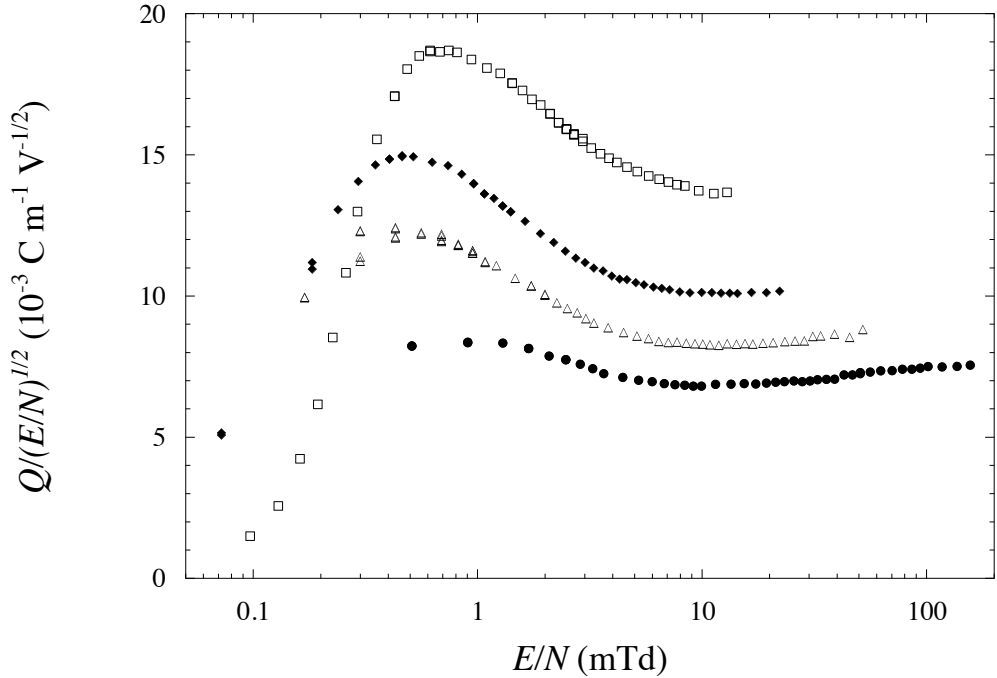}
\end{center}
\caption{\small Deviations of the experimental data from the YB law, $Q/\left(E/N\right)^{1/2},$ as a function of $E/N$ for $N(\mathrm{nm}^{-3})=3.09$  (squares),  $N=1.8 $ (diamonds),  $N=0.77$ (triangles), and $N=0.26$ (circles).\label{fig:QsuEsuN}}
 \end{figure}

In order to make deviations from the YB law more evident, we factor the $(E/N)^{1/2}$ dependence out of the data and plot $Q/(E/N)^{1/2}$ as a function of $E/N$ in figure \ref{fig:QsuEsuN} only for a few isopycnal curves again for the sake of clarity.
Except the smallest density for which very low values of $E/N$ were not reached, 
the deviations $Q/(E/N)^{1/2}$ are strongly peaked for all other densities in the range $(0.3<E/N<0.7)\,$mTd.

This behaviour of $Q/(E/N)^{1/2}$ very closely resembles that of the density--normalized electron mobility $\mu N$ as a function of $E/N$ (see figure 2 of reference \cite{BSL}). $\mu N$ shows a maximum that is related to the RT minimum of $\sigma_{\mathrm{mt}}.$  The similarity of the behaviour of $\mu N$ and $Q/(E/N)^{1/2}$ is hardly surprising because $\mu N $ is a suitable thermal average of $\sigma_{\mathrm{mt}}^{-1}$ \cite{BSL} and  $Q/(E/N)^{1/2}\propto \sigma^{-1/2}$ according to the YB model, equation \ref{eq:YBesun}.
The main difference between the behaviours of $\mu N$ and $Q/(E/N)^{1/2}$ is that the position of the mobility maximum occurs at a value $(E/N)_{m}$ that decreases with increasing $N,$ starting with $(E/N)_{m}\approx 4 \,$ mTd for $N=0.37\,$nm$^{-3}$ down to $(E/N)_{m}\approx 2\,$mTd for $N\approx 6.1\,$nm$^{-3}$ \cite{BSL}, whereas the position of the $Q/(E/N)^{1/2}$ maximum, though the quality of the data does not allow to locate it with great accuracy, apparently occurs for nearly the same values or for only slightly increasing values of $E/N$ for isopycnals of increasing $N.$ 

The decrease of $(E/N)_{m}$ with increasing $N$ in the mobility case has been rationalized \cite{BSL,Borghesani:2001uq} by realizing that the mobility maximum is the fingerprint of the RT minimum of $\sigma_{\mathrm{mt}}$ occurring for $\mathcal{E}\approx 230\,$meV. As the average energy of electrons is increased by the kinetic energy shift $E_{K}$ as $N$ increases, less energy $\propto E/N$ has to be supplied by the field with increasing $N$ in order that the average electron energy equals that of the RT minimum. 

Unfortunately, we do not have at present any similar, simple explanation for the behaviour of the $Q/(E/N)^{1/2}$ maximum. We can only argue that the values of $E/N$ corresponding to the maximum deviation from the YB law are smaller than those of the mobility maximum because in the present case the electrons are already epithermal upon injection.

Anyway, we want to show that the collected charge data still bear close relationship with the cross section even outside the range of (strict) validity of the YB model. In fact,
by using equation \ref{eq:YBesun} even at the point of maximum deviation of $Q$ from the $(E/N)^{1/2}$-law, one obtains
\begin{equation}
\sigma_\mathrm{Q,M}  \equiv \max{\left[Q/(E/N)^{1/2}\right]^{-2}}\propto \sigma
\label{eq:sigmaqmaxdev}\end{equation}
\begin{figure}[!t]
    \begin{center}
\includegraphics[width=\textwidth]{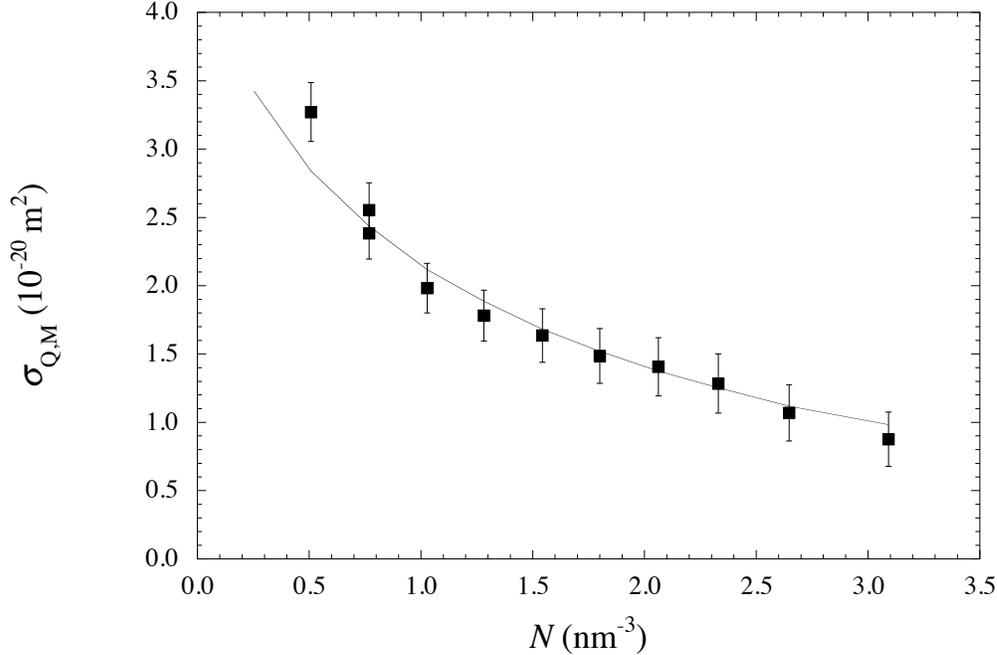}\end{center}
        \caption{\small  Cross section determined by using the maximum of $Q/(E/N)^{1/2}$ at low $E/N$ (squares). Solid line:  $ \sigma_{\mathrm{mt}}(\bar\mathcal{E})$ (see text). \label{fig:sigmaqm}}
 \end{figure}

\noindent In figure \ref{fig:sigmaqm} we plot  the values of $\sigma_{\mathrm{Q,M}}$ determined by using the same procedure as for equation \ref{eq:sigmasme}. The datum for the lowest density is not shown because a maximum is hardly observed at all for that density. $\sigma_\mathrm{Q,M}$ has thus been normalized to the average cross section $\sigma_{\mathrm{mt}}\left(\bar\mathcal{E}\right) $ calculated for $N=0.77\,$nm$^{-3}.$ $\sigma_{\mathrm{mt}}\left(\bar\mathcal{E}\right)$ is calculated theoretically from $\sigma_{mt}\left(\mathcal{E}\right)$ as explained before and, again, $\bar\mathcal{E}=(3/2)k_{\mathrm{B}}T +E_{K}(N).$

\noindent Once more, we note that the overall behaviour of the density dependence of the cross section determined at low $E/N$ is in excellent agreement with the model that takes into account multiple scattering effects. This fact validates the analysis carried out previously.

 \section{Conclusions} 

We have studied the collection efficiency of photoelectrons injected into dense argon gas at low temperature.  Our data cover a range of density--normalized electric fields $E/N$ much lower than previous data which were measured at higher fields by exploiting a different injection technique \cite{smejtek1973}.  
In the high--field region, our and previous data compare favorably with the only available theoretical model \cite{YB}. This model predicts that a fraction of the epithermal electrons injected into a gas may be returned to the cathode as soon as they undergo their first scattering event. According to this model, the momentum transfer scattering cross section can be deduced from the dependence of the collected charge on the reduced electric field. 

In view of the more complete, available knowledge about the scattering processes in dense rare gases that includes multiple scattering effects \cite{BSL}, we have been able to relate the cross section determined from the charge data to the thermal average of the gas--phase scattering cross section \cite{Weyhreter:1988kx}. 

Several problems, however, still remain unsolved. 
In our opinion, the most severe one deals with the hypothesis assumed by YB to derive their model. According to this hypothesis, the fate of an electron depends on its first encounter scattering. By contrast, we obtain a nice agreement with the model by calculating the thermal average of the gas--phase cross section and by taking into account multiple scattering effects. This fact means that electrons must have reached thermal equilibrium with the gas and it is very well known that thermalization occurs after a very large number of collisions \cite{kuntz:1136,Mozumder:1980uq}. This fact overtly contradicts the YB hypothesis. 

MC--based calculations, when showing that a very large number of electron--atom collisions are required to determine the 
fate of a given electron, confirm that an attempt at explaining the collection efficiency in terms of what occurs at the first encounter appears to be unrealistic (for a more complete discussion, see \cite{kuntz:1136}).

MC calculations do reproduce the observed $(E/N)^{1/2}$-law as a consequence of purely statistical effects \cite{kuntz:1136} though they do not suggest any physical explanations for the explicit square--root functional form. Moreover,
they fail at reproducing the density ordering of the experimental data in argon. This result is hardly surprising because MC studies have been carried out for classical trajectories without taking into account the quantum multiple scattering effects, which are active in a dense gaseous environment \cite{BSL}.

In any case, neither the YB model nor the MC calculations do provide an explanation of the change of behaviour of the collected charge as a function of the electric field at very low fields as we have observed.

 \section*{References}
 \bibliography{borghesanipsst}
\end{document}